\documentclass[3p,times,twocolumn]{elsarticle}

\usepackage{epsfig}
\usepackage{graphicx}
\usepackage{epstopdf}
\usepackage{bibentry}
\usepackage{amssymb}

\usepackage{lineno}

\usepackage[figuresright]{rotating}

\usepackage{amssymb}
\usepackage{subfigure}
\usepackage{setspace}

\begin{document}
	%
	\begin{frontmatter}
		
		\title{Characteristic study of a quadruple GEM detector and its comparison with a triple GEM detector}
		
		\author[a]{Rajendra~Nath~Patra}
		\author[a]{Rama Narayan Singaraju}
		\author[b]{Saikat Biswas}
		\author[a]{Yogendra P. Viyogi}
		\author[a,c]{Tapan K. Nayak}
		
		\address[a]{Variable Energy Cyclotron Centre, HBNI, Kolkata-700064, India}
		\address[b]{Bose Institute, Department of Physics and CAPSS, Kolkata-700091, India}
		\address[c]{CERN, Geneva~23, Switzerland}
		
		
		\begin{abstract}
			A quadruple GEM detector has been assembled in a standalone configuration and operated using Ar and CO$_2$ gas mixtures in proportions of 70:30 and 90:10. Detailed performance study of the detector has been made by using $^{106}$Ru-Rh $\beta$-source and X-ray spectrum of $^{55}$Fe source. Results of these measurements are presented in terms of gain, efficiency, energy and time resolutions and also compared with our earlier triple GEM results. The energy resolution has been found to be somewhat worse compared with that of the triple-GEM detector. Effect of drift field on electron transparency and time resolution has been studied in detail.
		\end{abstract}
		
	\end{frontmatter}
	
	
	\section{Introduction }\label{intro}
	
	In high energy physics (HEP) experiments, one of the essential steps in the extraction of 
	physical information from
	particle detectors are to accurately reconstruct the properties of the incoming particles. Multi-wire proportional counters have been traditionally used for this purpose in several large-scale experiments.
	The development of the Gas Electron Multiplier (GEM) detectors~\cite{sauli,sauli2016} over the last two decades with characteristics like better position resolutions, long-term stability, and high-rate handling,
	immediately attracted the attention of the HEP professionals, and attempts have been 
	made to integrate GEM based detectors into many experiments~\cite{altunbasCOMPASS,ketzer,benci,bozzoTOTEM}. 
	At the same time, several 
	studies have been made with different GEM detector configurations to achieve the requirements of the experiments. These have produced very promising results in terms of high rate radiation measurements, low discharge probability, low ion back flow (IBF), excellent position and energy resolutions as well as good time resolution~\cite{benci,alfonsi,ball, zhang}. GEM detectors are seen to be suitable for a long time stable operation without any appreciable aging in high radiation environment~\cite{alfonsi,altunbas,adak}.
	
	A GEM foil consists of an insulator made of a 
	50~$\mu$m thick Kapton foil with 5~$\mu$m thick copper cladding on both sides and pierced by a regular array of holes. 
	From the time of the invention of GEM lot of studies have been carried out with single GEM, double GEM, and triple GEM detectors. In a multi-stack GEM detector setup, the hole diameter, pitch, electric field 
	across different gas gap can be optimised to minimise the discharge probability~\cite{bachmannDis} and  IBF~\cite{ball,bachmannIbf,breskin,bondar,mormann}. It is also noticed that with increasing number of GEM foils the detector can be operated at a lower voltage across the GEM foils at a fixed gain~\cite{bachmannDis,breskin}. The advantages in multi-GEM detector like lower operating voltage, low discharge probability, and low IBF make it more attractive for large drift volume and high radiation environment.
	
	Over these years, most experiments have almost settled for triple GEM configuration for the detection of charged particles, as can be seen in both existing and planned future experiments~\cite{benci,bozzoTOTEM,cbmgem, tytgat,balla,arora,starFGT, adare}. Investigation of detectors with more than 3 GEM foils has been focused towards specific applications, like photon detection using CsI cathode \cite{vavra}, making a gaseous photomultiplier tube by Breskin's group~\cite{breskin,buzulutskov2000-1, buzulutskov2000-2}, in RICH application~\cite{blatnik} or where IBF reduction becomes essential as in the ALICE Time Projection Chamber (TPC)~\cite{alicetpc}.
	The quadruple GEM configuration used for ALICE TPC readout is very special employing different hole size and pitch in different layers.
	
	In the present study, 
	we have constructed a quadruple GEM detector and performed a detailed characteristic study in the standalone configuration. 
	Here we present the results of our finding in terms of 
	the effective gain, energy resolution and efficiency measurements with Ar/CO$_{2}$ 90:10 and 70:30 gas mixtures. The results obtained in this study will also be compared with our triple GEM results~\cite{rajendra2017, rajendra2017IEEE}. The effect of the drift field (E$_{d}$) on the gain variations and time resolution are also studied in detail. 

	\section{Detector construction and test setup}
	
\begin{figure}[!th]
		\centering
		\includegraphics[width=0.45\textwidth]{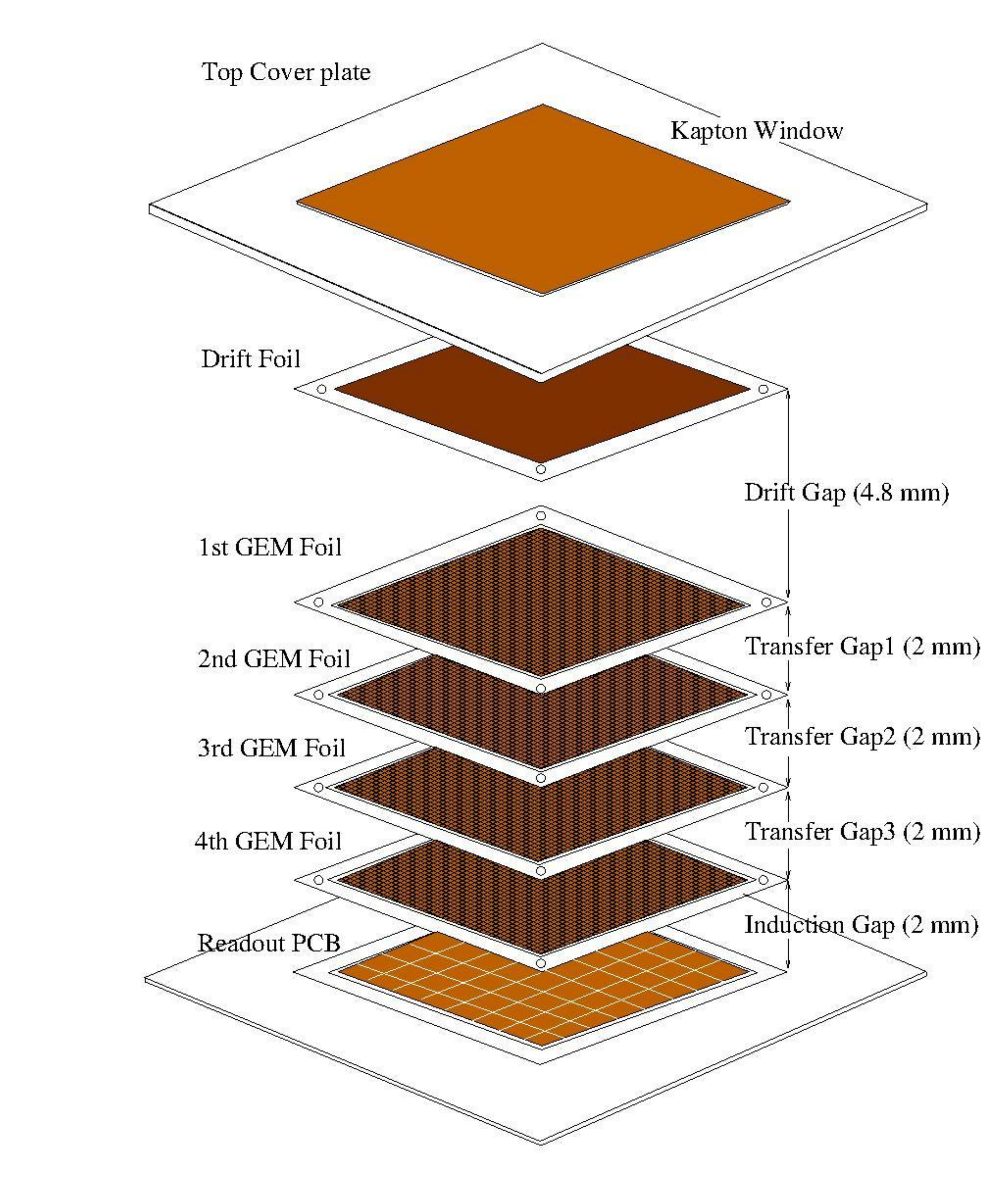}
		\caption{Schematic design of a quadruple GEM detector.}
		\label{schematic}
	\end{figure}

	A quadruple GEM detector has been assembled for detailed characteristics study at Variable Energy Cyclotron
	Centre, Kolkata, India. The components used in the detector assembly are obtained from CERN, Geneva. The GEM foils, used in this detector, have 70~$\mu$m diameter hole and 140~$\mu$m pitch in a hexagonal pattern with 10$\times$10~cm$^{2}$ active area. The position of the GEM foils and the different gaps of the detector are shown in a schematic diagram in Fig.~\ref{schematic}. Four GEM foils are placed one above the other followed by a drift (cathode) plane on the top, above the readout PCB. The readout plane has 120~pads of an equal area over the 10$\times$10~cm$^{2}$  PCB surface for picking up signals. The drift plane, GEM foils, and the readout plane are covered by G10 frame with a window on the top covered by a Kapton foil for the gas tightness. The drift gap, transfer gaps and the induction gap of the detector are 4.8-2-2-2-2~mm, respectively.
	The present configuration is somewhat similar to that reported recently~\cite{swain}, but very different from the one employed in the ALICE TPC~\cite{alicetpc} and also in other multi-GEM detectors reported in literature~\cite{breskin,blatnik}.
	
	For the construction of the detector, at first all the GEM foils are passed through quality assurance (QA) tests. In the QA testing procedure, all the GEM foils are kept in a Nitrogen environment and 500~V is applied across the foils. The leakage currents of the GEM foils are measured for 30 minutes. The acceptance criteria of the foils are that the current must be below
	500~pA without any major spark. A quadruple GEM detector is assembled as per the schematic of Fig.~\ref{schematic} with the foils which have passed the QA.
	The detector is operated with Ar and CO$_2$ gas mixtures with two different ratios, 90:10 and 70:30 at atmospheric pressure. 
	\begin{figure}[!th]
		\centering
		\includegraphics[width=0.45\textwidth]{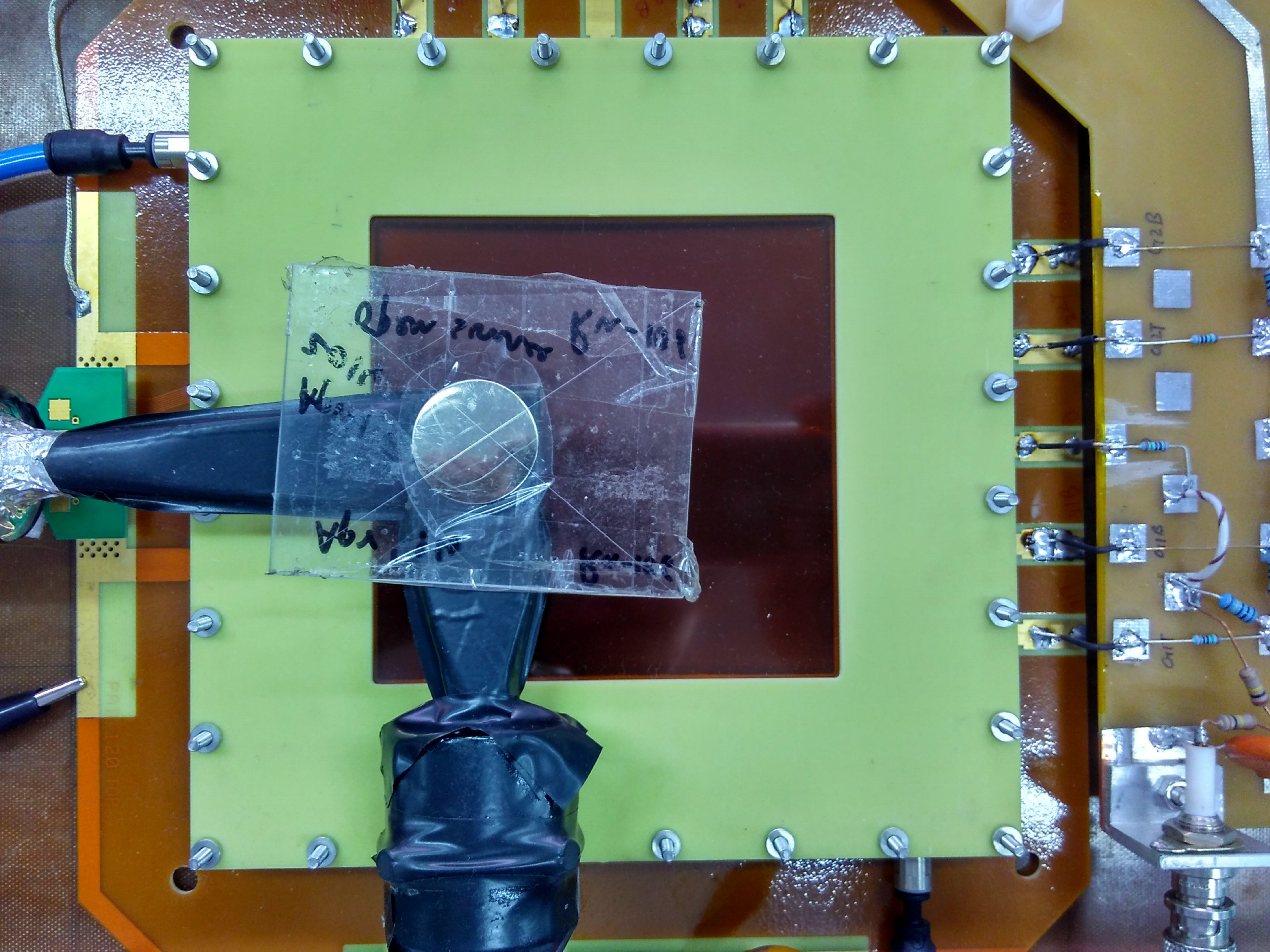}
		\caption{A photograph of the laboratory setup of the quadruple GEM detector.}
		\label{setup}
	\end{figure}
	\begin{figure}[!th]
		\centering
		\includegraphics[width=0.45\textwidth]{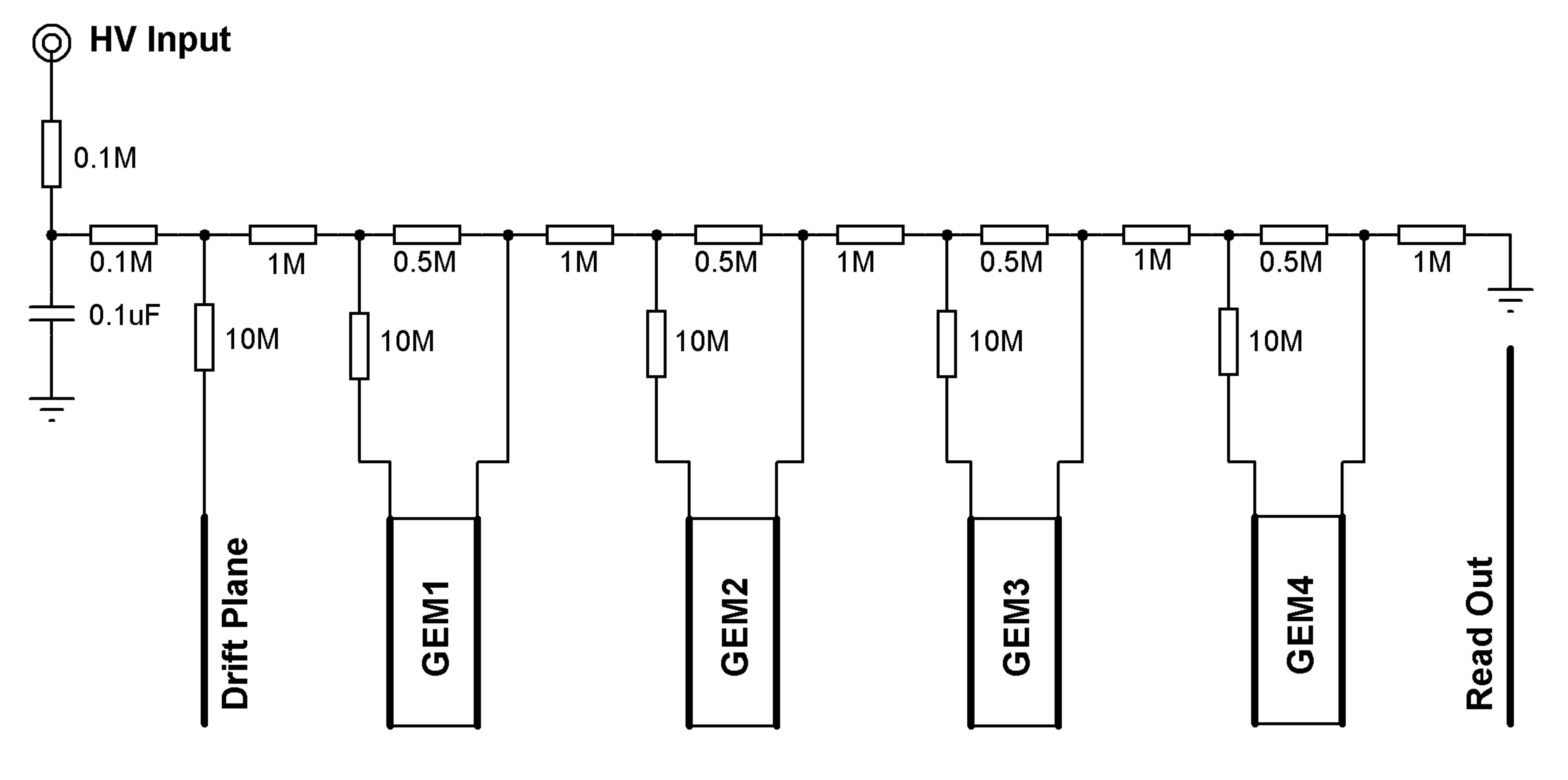} \\
		\includegraphics[width=0.45\textwidth]{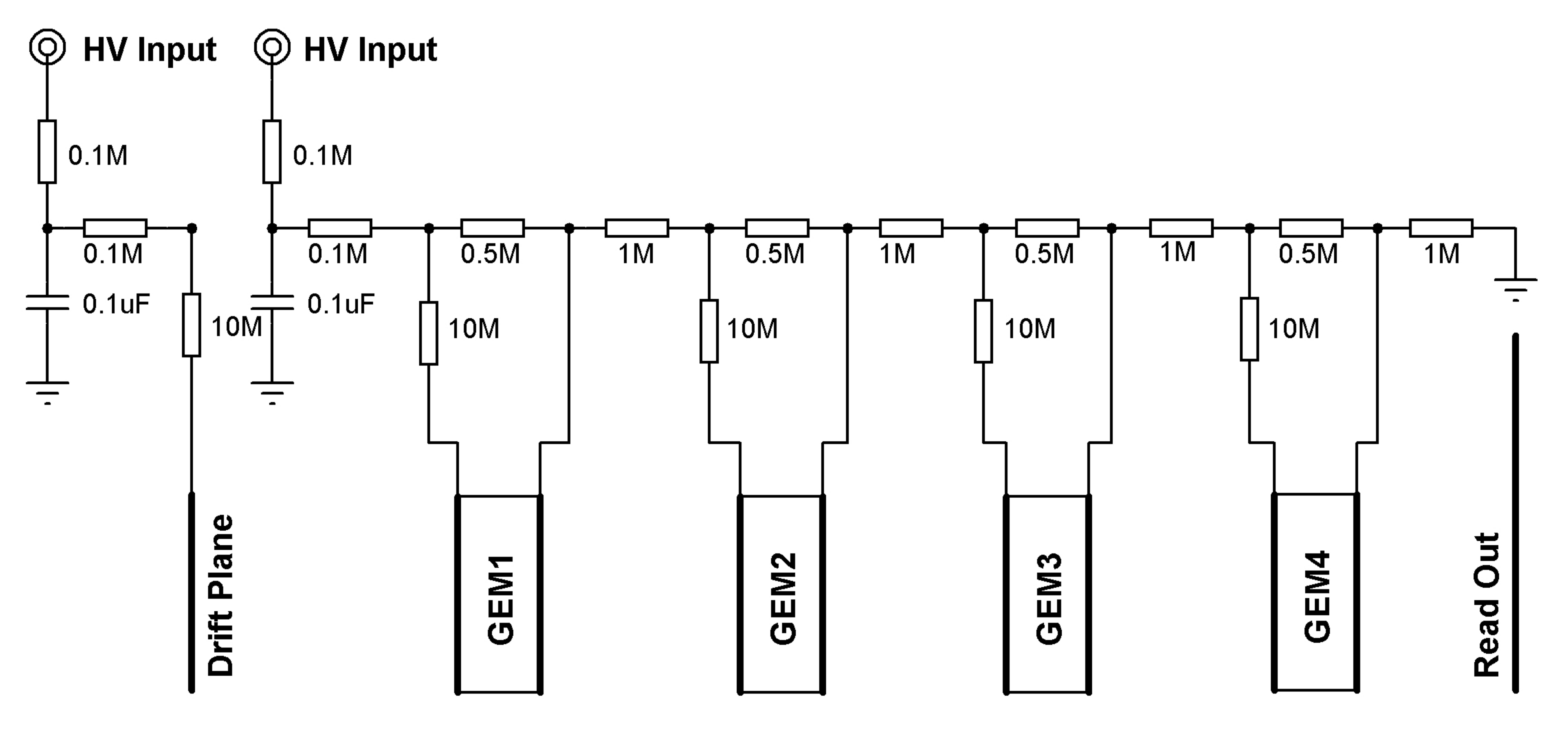}
		\caption{Top panel shows the 
			resistor chain used for voltage divider in normal configuration of the quadruple GEM
			during the tests. Bottom panel shows the resistor chain for voltage divider for 
			independent drift field (E$_{d}$) configuration.}
		\label{resistance_chain}
	\end{figure}
	
	The experimental setup of the detector is shown in Fig.~\ref{setup}. The high voltage (HV) is applied through a resistor chain to produce required electric field within the holes of the GEM foils and in the different gaps. The resulting electric field within the tiny holes is very high, $\sim$60~kV/cm so that it can produce charge avalanches. The field in different gaps is $\sim$1-2~kV/cm only for the drift of electrons. The details of the resistor chain used in this study are shown in Fig.~\ref{resistance_chain}. The resistor chain with single HV channel as in the top panel of  Fig.~\ref{resistance_chain} is used for the measurements of Section~\ref{sec_eff1}$-$\ref{sec_eff}. 
	In a second scheme, two HV channels are used as shown in the bottom panel of
	Fig.~\ref{resistance_chain}. This is used to change the drift field independently without changing any other field configurations. This scheme is used for the study of the effect of drift field on electron transparency (see Section~\ref{sec_drift}) and on the time resolution (see Section~\ref{sec_time_reso}) of the detector.
	
	In this test setup, NIM (Nuclear Instrumentation Module) based electronics modules have been used for the signal processing. A sum-up board is used to add up signals from all the 120 readout pads. 
	The summed-up signal is processed by a charge sensitive pre-amplifier (model ORTEC 142IH) which converts the charge signal to voltage signal. The noise level is $\sim$1100 electrons and the sensitivity is 1~mV/fC.
	The capacitance between the detector and pre-amplifier has been obtained to be 68~pF.
	The output voltage signal is fed to an amplifier (model ORTEC 572) for all the measurements, except for the measurement of time resolution where a fast amplifier is used as discussed in Section~\ref{sec_time_reso}. The signal shaping time of the amplifier is set to 1~$\mu$s 
	to collect maximum charges of the signal and the radioactive sources used here have low rate so no pile-up of the signal is expected with 1~$\mu$s shaping time. The output voltage signal of the amplifier is fed to a 12-bit ADC (model ORTEC ASPEC-927) with resolution 2.44~mV/ADC channel. The ADC module acquires a single spectrum for voltages without any discrimination. The electronic modules used in this measurement have been calibrated to avoid any errors.
	It is to be noted that no external threshold has been applied in the electronics readout. 
	However, the ADC module was insensitive for voltage \textless~40~mV, so there is no count at low ADC channel in the energy spectrum.
	\section{Test results and discussion}
\begin{figure}[!th]
		\centering
		\includegraphics[width=3.0in, height=2.0in]{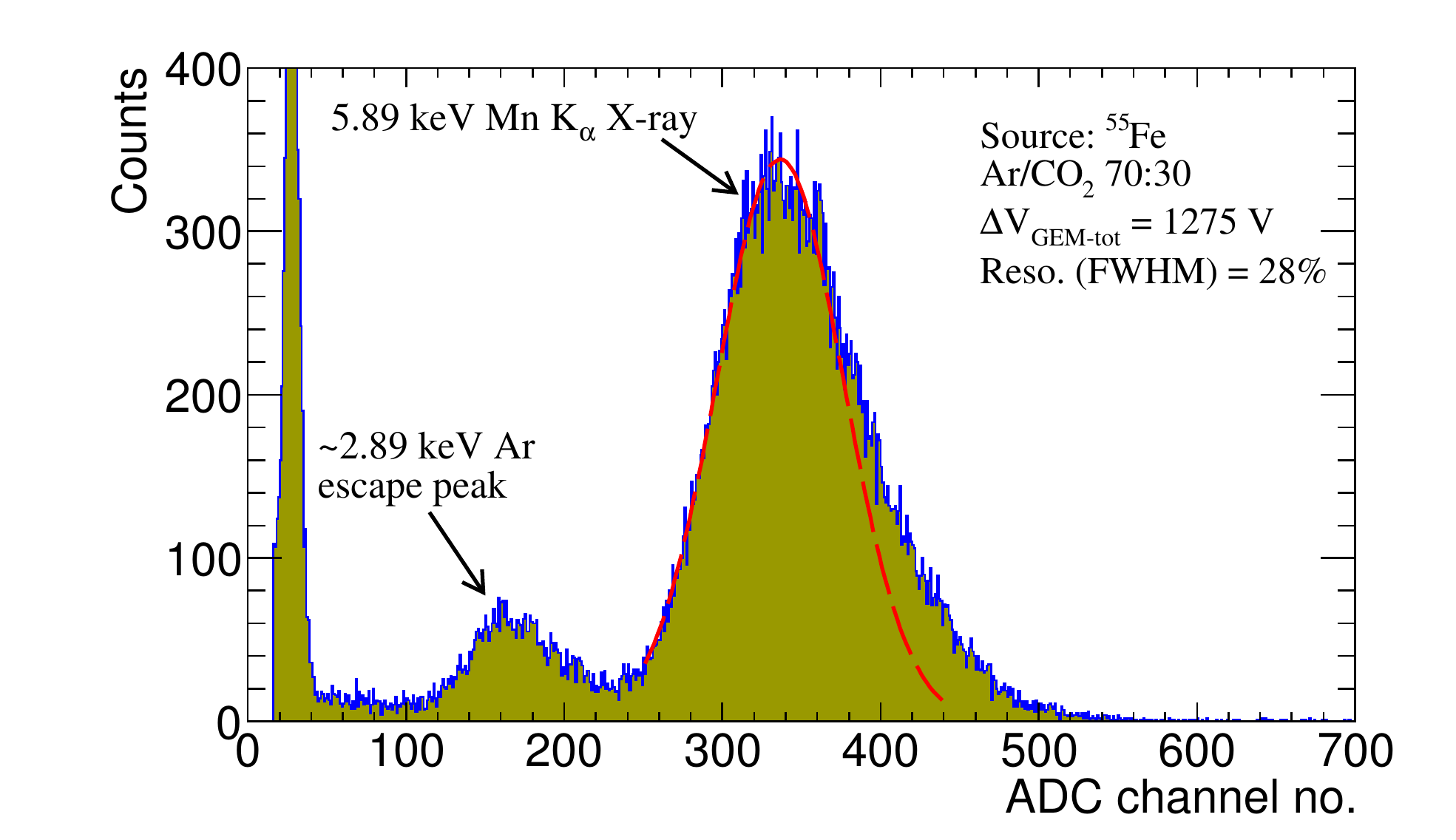}
		\caption{The pulse-height spectrum of $^{55}$Fe X-ray source of the quadruple GEM detector. 
			The peak at $\sim$2.89~keV corresponds to the Ar escape peak. 
			The main photo peak corresponds to the K$_{\alpha}$ X-ray of energy 5.89~keV.}
		\label{spectrum-fe}
	\end{figure}

	Tests of the detector have been carried out using radioactive sources after stabilisation of the gas flow
	and minimisation of the electronic noise.
	Figure~\ref{spectrum-fe} shows the $^{55}$Fe X-ray spectrum measured by the quadruple GEM detector set up at the
	total GEM voltage,  $\Delta V_{GEM-tot}=1275~V$ for Ar/CO$_{2}$ of 70:30 gas mixture.
	$^{55}$Fe decays via electron capture producing  $^{55}$Mn with a half-life of 2.73 years. The vacancy in the K-shell produced by electron capture is shortly filled by an electron from a higher shell. In this process, K$_{\alpha}$ X-ray of energy 5.89~keV is emitted. 	
	The smaller peak corresponding to $\sim$2.89~keV is the Ar K-shell X-ray escape from the gas volume known as Ar escape peak. 
	The main photopeak of $^{55}$Fe source of energy 5.89~keV along with Ar escape peak is shown in Fig.~\ref{spectrum-fe}.  An asymmetric broadening of the peak is observed, which could arise because of detector effects.
	The $^{55}$Fe spectra at different gas mixtures and voltages
	have been analyzed in detail to calculate effective gain, energy resolution, and electron transparency.

	\begin{figure}[!th]
		\centering
		\includegraphics[width=3.0in, height=2.0in]{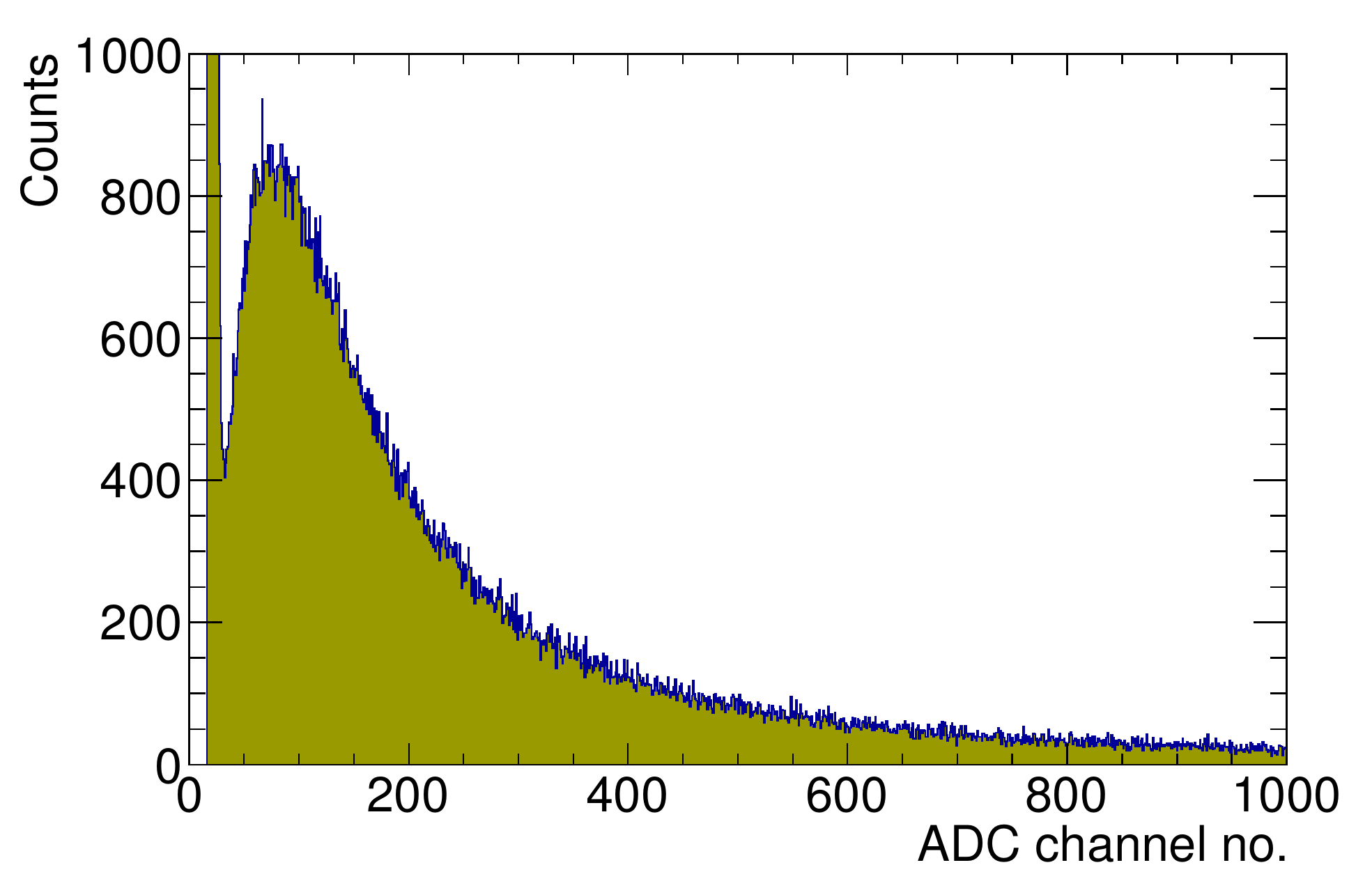}
		\caption{The pulse-height spectrum of $^{106}$Ru-Rh $\beta^{-}$-spectrum at $\Delta V_{GEM-tot}$ 1316~V in Ar/CO$_{2}$ 70:30 gas.}
		\label{spectrum-ru}
	\end{figure}
	
	The next set of tests have been performed by exposing the GEM detector to $^{106}$Ru-Rh $\beta^{-}$
	source. The resulting spectrum, taken at $\Delta V_{GEM-tot}=1316$~V
	in Ar/CO$_{2}$ 70:30 gas mixture, is shown in Fig.~\ref{spectrum-ru}. The $\beta$-spectrum is continuous
	with a long tail extending to large ADC values. The $^{106}$Ru-Rh source has also been used for the efficiency and
	time resolution measurements.

	\subsection{Detector efficiency and operating condition}\label{sec_eff1}
\begin{figure}[!th]
		\centering 
		\includegraphics[width=3in]{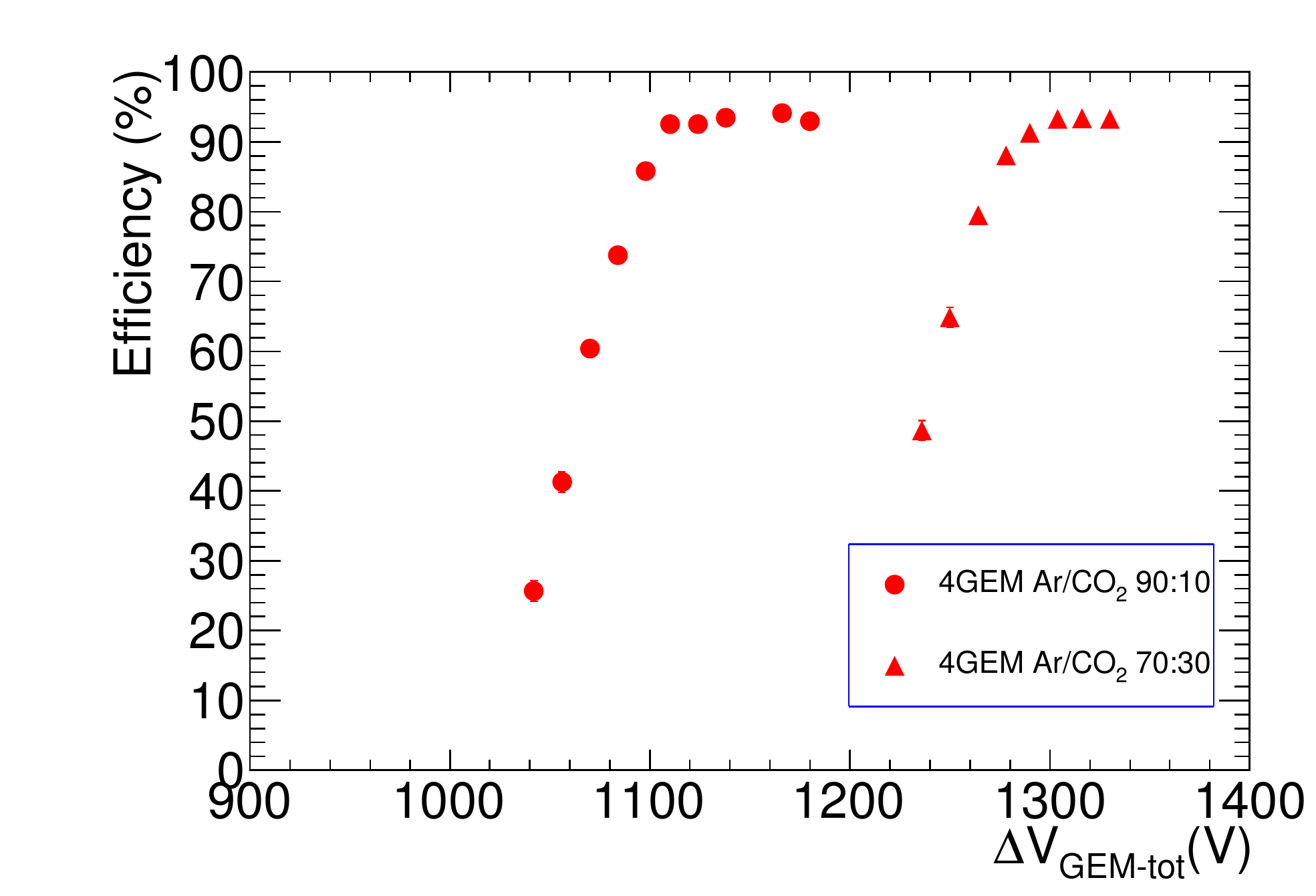}
		\caption{Efficiency of the quadrupole detector as a
			function of $\Delta V_{GEM-tot}$ is presented
			for two Ar/CO$_{2}$ gas mixtures.  The statistical errors
			are small (within a couple of percent) and within the symbol size.}
		\label{efficiency}
	\end{figure}

	Measurement of the efficiency for the quadruple GEM detector is done in a similar manner to the
	efficiency measurement of the triple GEM~\cite{rajendra2017}. The
	trigger was provided by the coincidence signal of
	a set of three detectors, two cross scintillators placed above the detector and a third scintillator placed below.
	All the three scintillators are placed in a configuration to maximise the overlap. 
	The $^{106}$Ru-Rh $\beta^{-}$-source is placed on the top of 
	the cross scintillators. The scintillators are chosen to be very thin ($\sim$2~mm) so that
	$\beta^{-}$ particles penetrate through those. The number of triggered particles giving the signal on the GEM detector yields the efficiency. 
	The variation of efficiency as a function of applied GEM voltage for the two gas mixtures
	is shown in Fig.~\ref{efficiency}. Efficiency rapidly increases with
	GEM voltage and then saturates after certain voltage.
	The starting of the plateau region corresponds to the operational voltage of the GEM detector, 
	which in the present case are 1120~V and 1300~V for the 90:10 and 70:30 gas mixtures, respectively.
	The efficiency at the plateau is $\sim$94\%. The absolute value of the
	efficiency depends on the electronics and trigger particle. The ADC module used in this measurement was insensitive below 40 mV. This may result in some loss of efficiency and thus the efficiency does not reach 100\%.

	\subsection{Detector gain}\label{sec_gain}
\begin{figure}[!th]
		\centering 
		\includegraphics[width=0.45\textwidth]{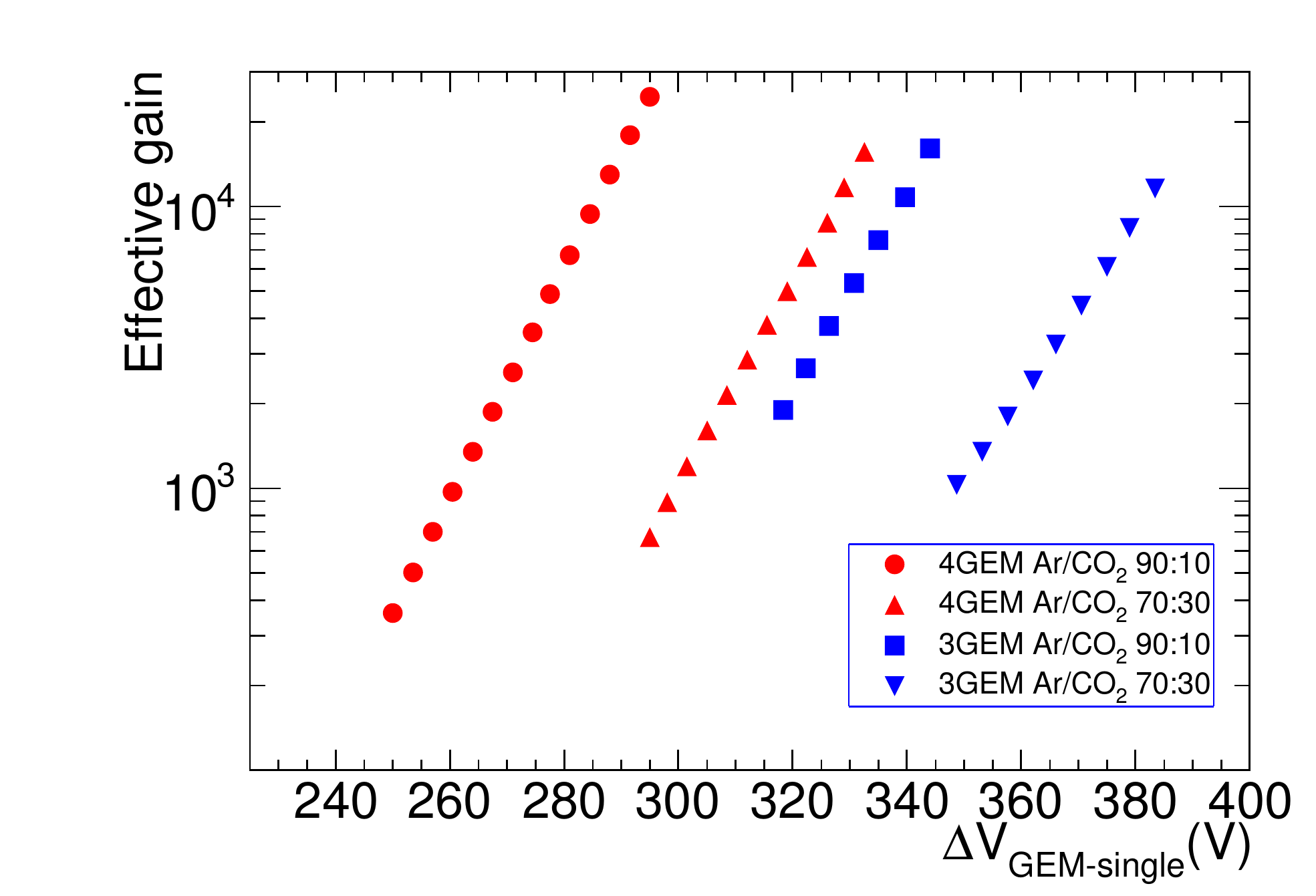}
		\caption{Effective gain as a function of $\Delta
			V_{GEM-single}$ is presented with different Ar/CO$_{2}$ gas mixtures for the quadruple
			GEM and triple GEM detectors. 
			The statistical errors are small and within the symbol size.}
		\label{gain}
	\end{figure}

	GEM detector gain has been calculated from the 
	main photo peak of the $^{55}$Fe 5.89~keV X-ray spectrum. 
	In Fig.~\ref{spectrum-fe}, we have shown the Gaussian fit to the spectrum, from where
	we obtain the peak position and width corresponding to 5.89~keV energy.  
	Gain values are calculated according to the procedure given in Ref.~\cite{rajendra2017}.
	Gains are plotted as a function of the GEM voltage for both the gas mixtures, as shown in 
	Fig.~\ref{gain}. For a given value of the GEM voltage ($\Delta V_{GEM-single}$), the gain for 90:10 gas mixture is much larger compared to that of the 70:30 gas mixture. For comparison, results of the triple GEM detector~\cite{rajendra2017} are also included in the figure. 
	It is important to notice that to have similar gain the individual GEM voltage required in case of triple GEM detector is higher than that for the quadruple GEM detector.

	The GEM foils operate at the low voltage across it have low gas gain
	which corresponds to stable operation against discharge within the GEM. Because
	discharge depends on the gain and it is proportional to the gain
	value. We have not measured the stability of GEM against discharge in
	this study. However, the discharge study of GEM detector is reported
	in past by Bachmann et al.~\cite{bachmannDis} for multiple GEM
	configuration and also discussed in Ref.~\cite{alicetpcADD1}. They have shown that the detector is less prone to discharge with increasing number of GEM foils and it also depends on the voltages applied across the foils.
	\subsection{Energy resolution of the detector}\label{sec_reso}
\begin{figure}[!th]
		\centering 
		\includegraphics[width=0.45\textwidth]{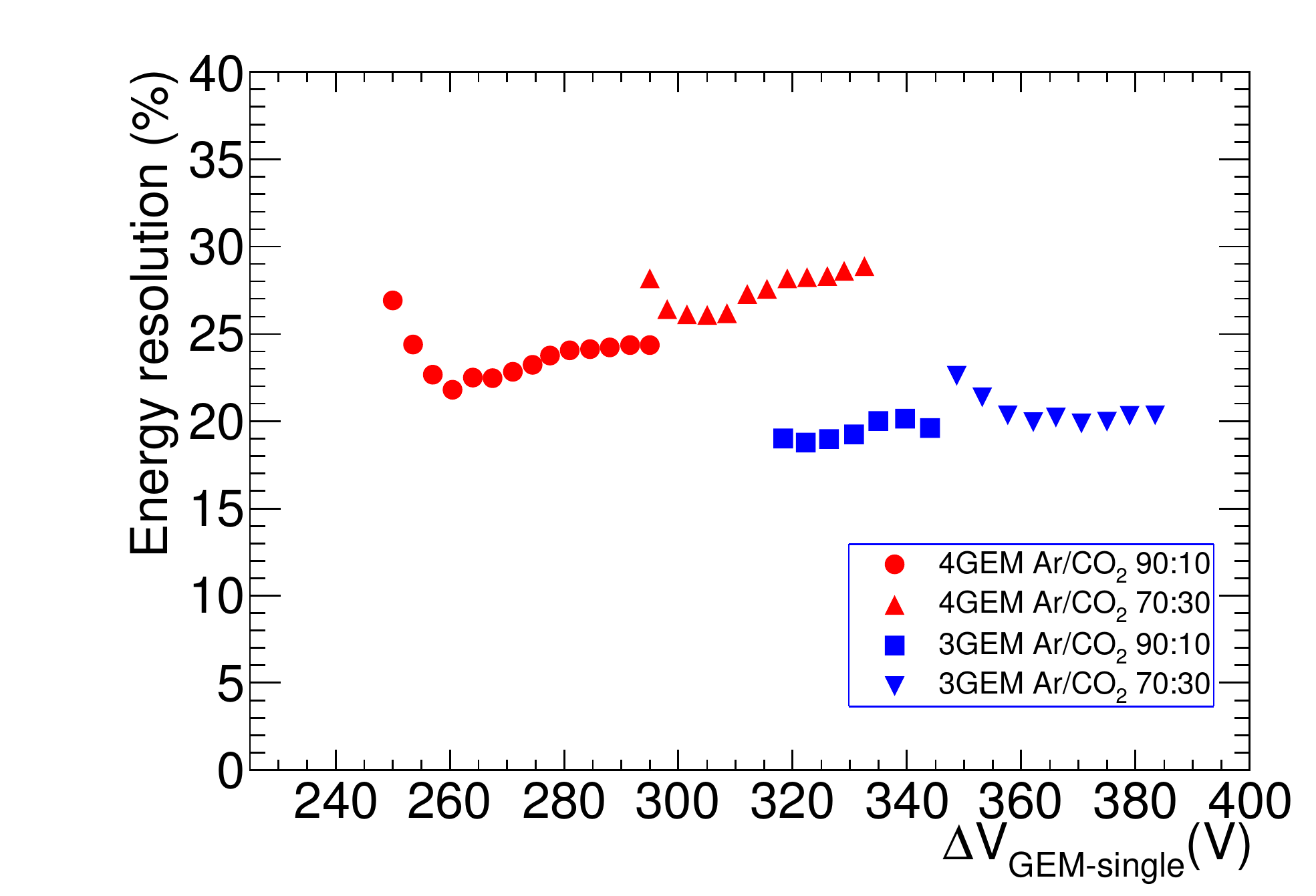}
		\caption{ Energy resolution as a function of $\Delta
			V_{GEM-single}$ is presented with 
			different Ar/CO$_{2}$ gas mixtures for both quadrupole and
			triple GEM detectors. The statistical errors are small and within the symbol size.}
		\label{reso}
	\end{figure}

	The energy resolution is calculated from the Gaussian fit parameters of the 5.89~keV peak of the $^{55}$Fe spectrum. 
	The energy resolution in terms of full width at half maximum (FWHM) 
	as a function of the individual GEM voltage is shown in
	Fig.~\ref{reso} for two different gas mixtures in case of both  
	quadruple GEM and triple GEM configurations. Energy resolutions are
	somewhat better for 90:10 gas mixture compared to 70:30. For both gas
	mixtures, the energy resolutions in case of quadruple GEM detector are
	larger compared to those of the triple GEM detector. Similar results
	have also been reported in a previous study~\cite{alicetpc}, where the
	effect has been attributed to low IBF in case of quadruple GEM detector. 
	Low IBF is achieved in the quadruple GEM detector in low field
	configuration so that ions are stopped at the copper part of the
	GEMs. The field setting affects the electron transmission
	efficiency. In particular, low voltage in the first GEM favours low
	IBF. However, it causes the losses of the primary electrons from the
	drift volume and low gain as a result of worsens the energy resolution.
	\subsection{Comparison of efficiency for triple GEM and quadruple GEM detectors} \label{sec_eff}
\begin{figure}[!th]
		\centering 
		\includegraphics[width=2.7in]{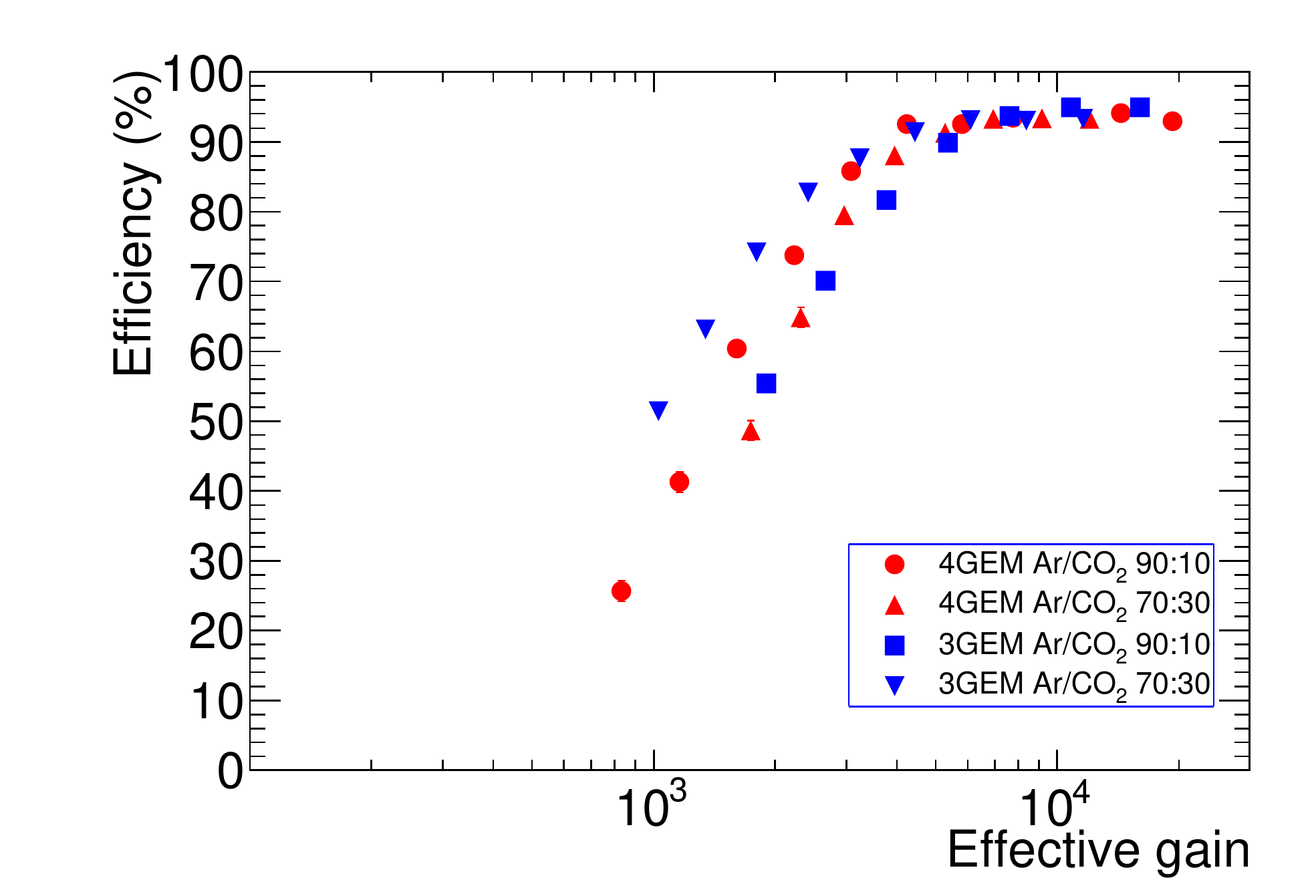}
		\caption{Variation of efficiency as a function of effective
			gain with different Ar/CO$_{2}$ gas mixtures for both triple
			GEM and quadruple GEM detectors. The statistical errors are small and within the symbol size.}
		\label{efficiency-gain}
	\end{figure}

	The magnitude of the detector signal depends on the 
	amount of charge collected by the readout pads and hence 
	on the factors like gain, number of primary ionization and sensitivity
	of the electronics. 
	For fully absorbed $^{55}$Fe X-ray the number of primary ionization produced within the drift volume depends on the gas mixture. Same readout electronics have been used
	for the study of both the triple GEM and quadruple GEM detectors. Thus the gain
	should be the only factor for efficiency determination irrespective of
	the kind of the detector and gas mixtures used in the study.
	For a given applied $\Delta V_{GEM-single}$, the gain of the detectors have been measured using
	$^{55}$Fe source and the data have already been shown in Fig.~\ref{gain}. For the
	same value of $\Delta V_{GEM-single}$ efficiency value can be taken
	from Fig.~\ref{efficiency}. Similarly, data for triple GEM detector
	have been taken from the Ref.~\cite{rajendra2017}. A compilation of
	the two data sets for efficiency as a function of
	effective gain has been presented in Fig.~\ref{efficiency-gain}. It is observed that efficiency
	increases with effective gain and then comes to a plateau after
	certain gain. It is to be noted that the efficiency plateau starts from a similar gain value in all four cases, irrespective of the
	kind of GEM detector and its gas compositions. Thus gain plays a principal
	role in efficiency measurement. It is found that
	efficiency reaches a maximum value at the gain $\sim$5000 for both
	the detectors.
	\subsection{Drift field effect and electron transparency } \label{sec_drift}
	
	Drift electric field (E$_{d}$) of a GEM detector is defined as the field between 
	drift plane (cathode) and the top of the first GEM foil. In the presence of drift field, the primary electrons produced within the drift gap drift towards the top GEM foil and transfer through the holes for the electron multiplication and go through subsequent steps. The number of primary electrons transferred through the holes of the top GEM foils depends on the drift electric field and the top GEM field~\cite{bachmannIbf}. 
	
	The effect of drift field (E$_{d}$) in the electron transparency was studied for the quadruple GEM detector for both the gas mixtures using $^{55}$Fe. In this setup, the resistor
	chain is shown in the bottom panel of Fig.~\ref{resistance_chain} was
	used to change drift field and GEM voltages independently. The mean value of the ADC channel no. has been calculated from the 5.89~keV
	X-ray peak at different field values keeping GEM voltages constant. The total charge collection on the
	readout depends on electron multiplication and hence on GEM voltages,
	therefore the results with different GEM voltage sets are normalized
	to unity for comparison with the assumption that full transparency is
	reached at the plateau region. 
	The electron transparency of the detector is defined as
	the ratio of primary electrons collected in the
	first GEM holes and the number of primary electrons
	created within the drift volume.
	
	\begin{figure}[!th]
		\centering 
		\includegraphics[width=0.45\textwidth]{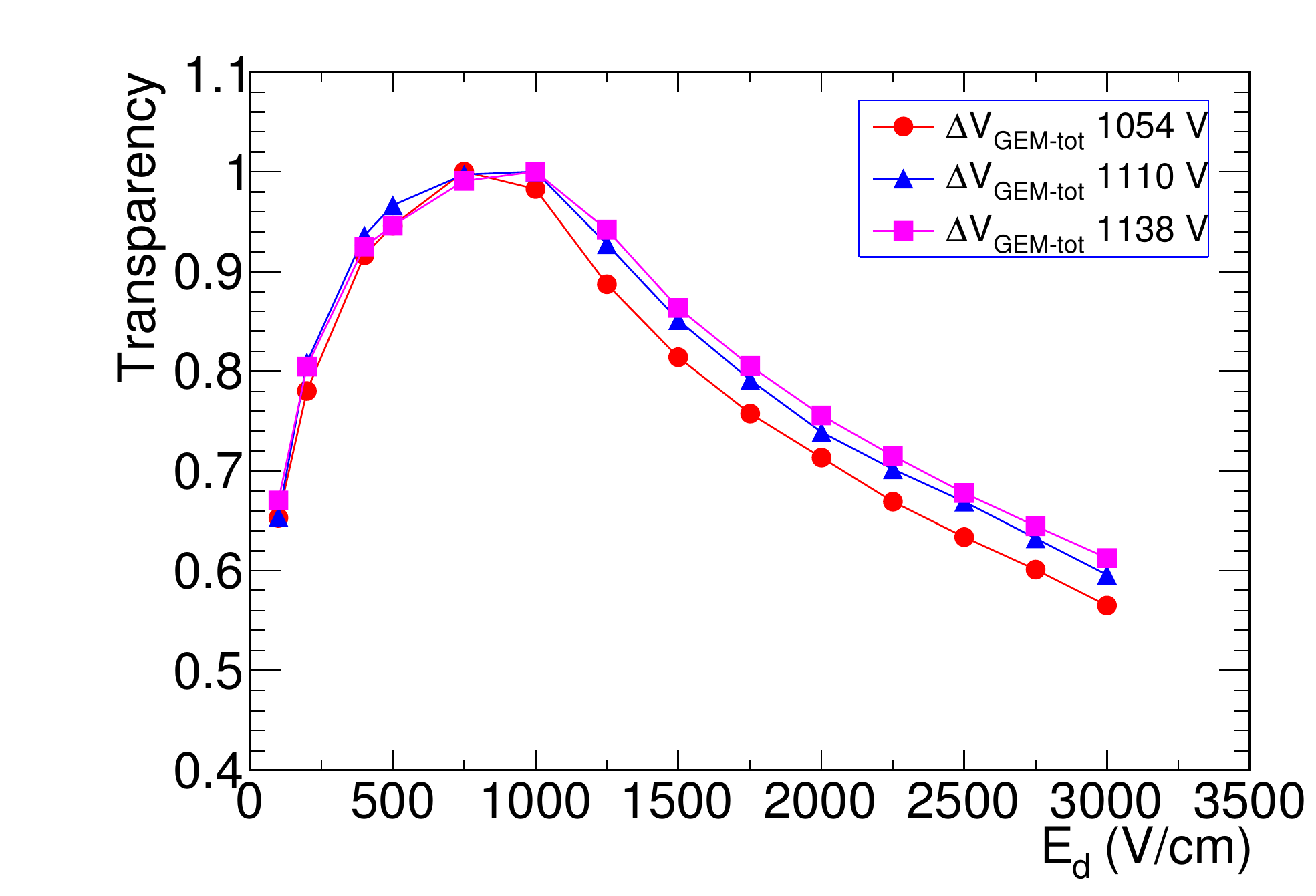}
		\includegraphics[width=0.45\textwidth]{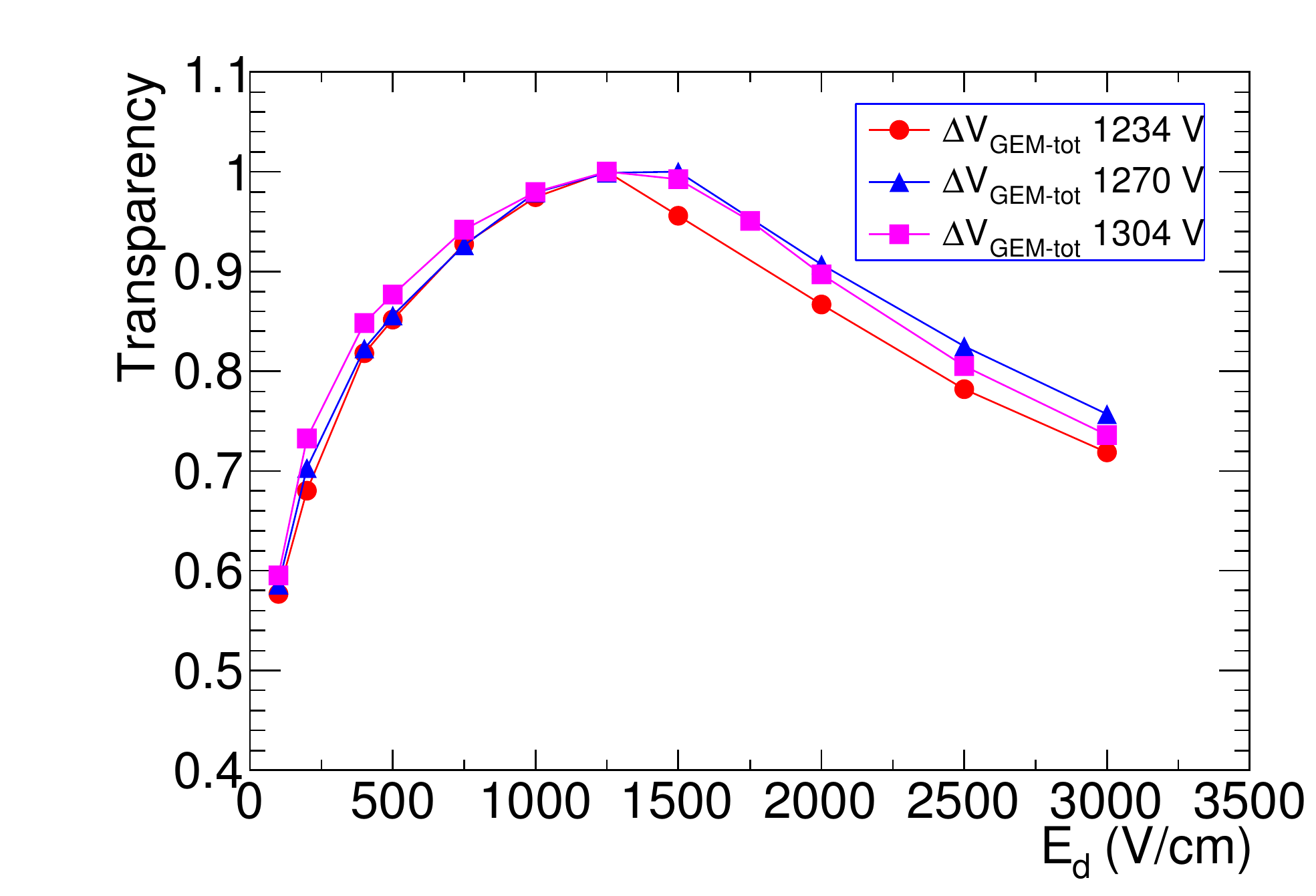}
		\caption{Electron transparency as a function of drift field (E$_{d}$). (top) Ar/CO$_{2}$ 90:10 and (bottom) Ar/CO$_{2}$ 70:30 gas mixtures with different $\Delta V_{GEM-tot}$ sets. The statistical errors are small and within the symbol size.}
		\label{drift_transparency}
	\end{figure}
	
	Figure~\ref{drift_transparency} shows the electron transparency as a
	function of E$_{d}$ for Ar/CO$_{2}$ with 90:10 (upper panel) and 70:30
	(bottom panel) gas mixtures. The values of transparency increase with
	E$_d$ attains an optimum value and then decreases. 
	The optimum transparency of the electrons for the Ar/CO$_{2}$ 90:10
	and 70:30 gas mixtures are found at drift field (E$_{d}$) values of
	750-1000~V/cm and 1000-1500~V/cm, respectively.
	Earlier it was seen that the operating GEM voltage of Ar/CO$_{2}$ 70:30 gas
	mixture is higher compared to that of the Ar/CO$_{2}$ 90:10 gas
	mixture. The drift fields
	within the holes of Ar/CO$_{2}$ 70:30 are higher for in comparison to
	the 90:10 gas mixture, and so the drift electrons could be focused
	through it more efficiently. 
	At higher  E$_{d}$, the electron transparencies of Ar/CO$_{2}$ 70:30
	remain relatively high.
	
	The results are in qualitative agreement with similar studies for single
	GEM carried out by S.~Bachmann~\textit{et al.}~\cite{bachmannIbf}. The
	electron collection efficiency or transparency increases initially,
	attain a plateau and decreases after that. Low collection
	efficiency at lower field value might occur because of losses of
	electrons due to diffusion, recombination or attachment of the
	electrons with the contaminant of the gas. However, at higher drift
	field the electric field lines are terminated at the top conducting
	surface of the GEM. Some ionization electrons possibly might be
	collected on the GEM surface.

	\begin{figure}[!th]
		\centering 
		\includegraphics[width=2.7in]{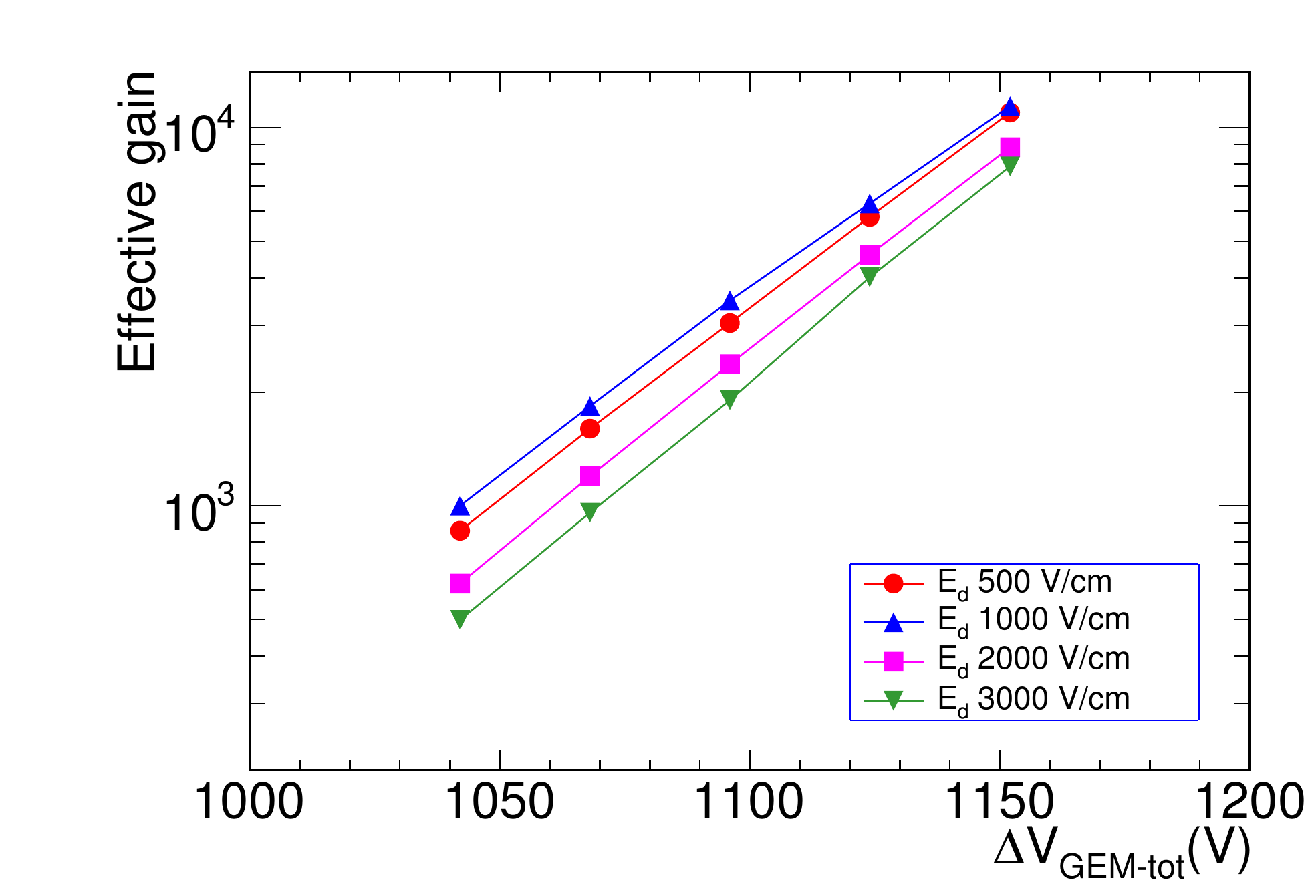}
		\includegraphics[width=2.7in]{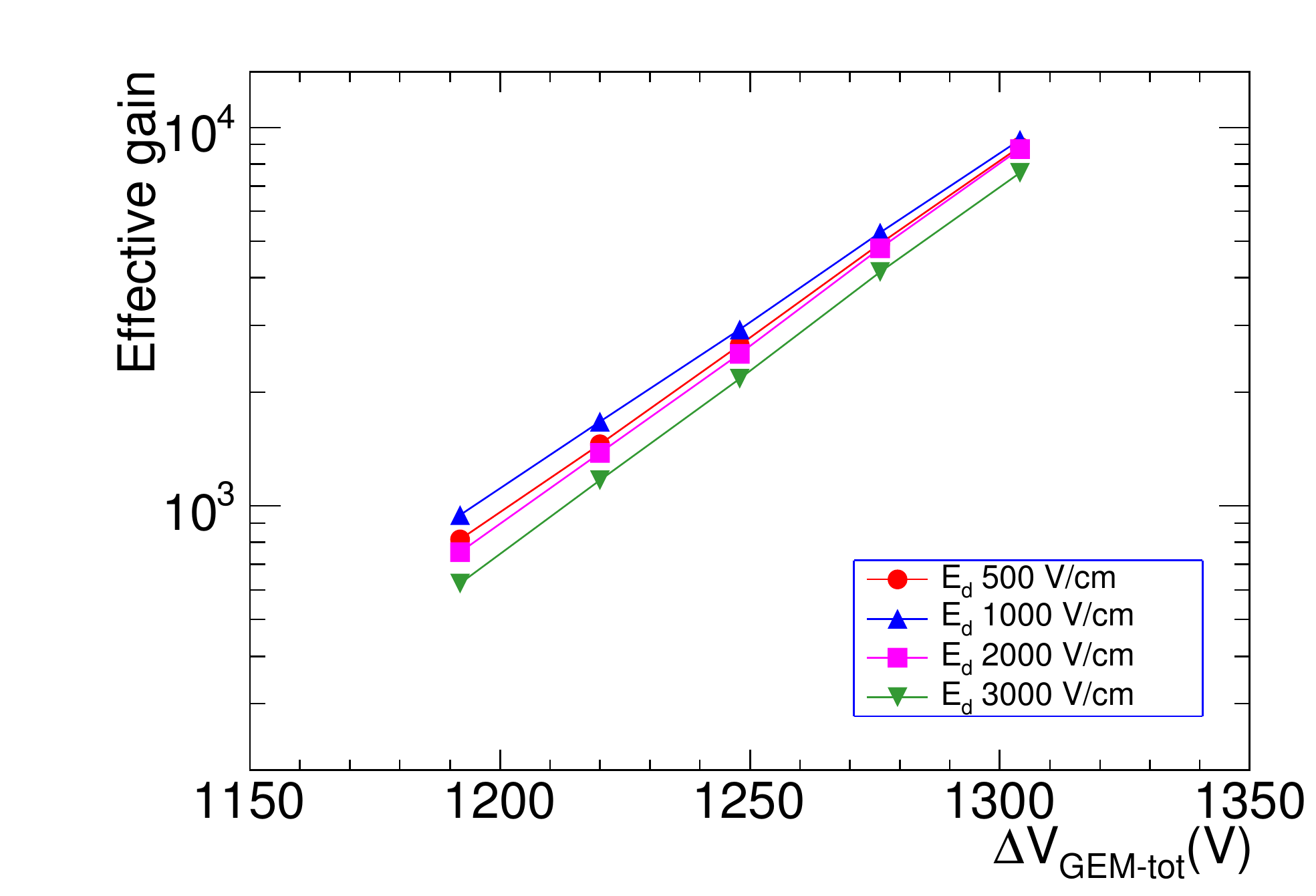}
		\caption{Gain variation as a function $\Delta V_{GEM-tot}$ is given for (top) Ar/CO$_{2}$ 90:10 and (bottom) Ar/CO$_{2}$ 70:30 gas mixtures with different drift field (E$_{d}$) sets. The statistical errors are small and within the symbol size.}
		\label{drift_gain}
	\end{figure}

	The effect of drift field on the operation of GEM has been further
	explored. The effective gain of the detector is also measured by
	varying the GEM voltages with different drift field (E$_{d}$)
	sets. The results are shown in Fig.~\ref{drift_gain}. It can be found
	that the highest gain corresponds to E$_{d}$$\sim$1000~V/cm, which is
	the closest drift field for the maximum electron transparency for both
	the gas mixtures. The somewhat larger differences in gain values at
	lower $\Delta V_{GEM-tot}$ are found to narrow down with increasing
	$\Delta V_{GEM-tot}$.
	The reason for this is that the transparency depends on both first GEM
	and the drift field. At lower GEM voltages, drift field effect is quite
	strong as was seen in the gain value. However, at higher GEM voltages, the electric field across GEM is so large that a significant fraction of the
	electrons are focused through the hole. Hence, electron transparency
	increases. So the drift field becomes less important and the gain
	values converge to a single value.  From
	Fig.~\ref{drift_transparency} and \ref{drift_gain} it can be concluded
	that both drift field and field across the holes of the first GEM have
	important roles in electron transparency depending on the field range.

	\subsection{Time resolution}
	\label{sec_time_reso}
	
	\begin{figure}[!th]
		\centering 
		\includegraphics[width=3.0in, height=2.0in]{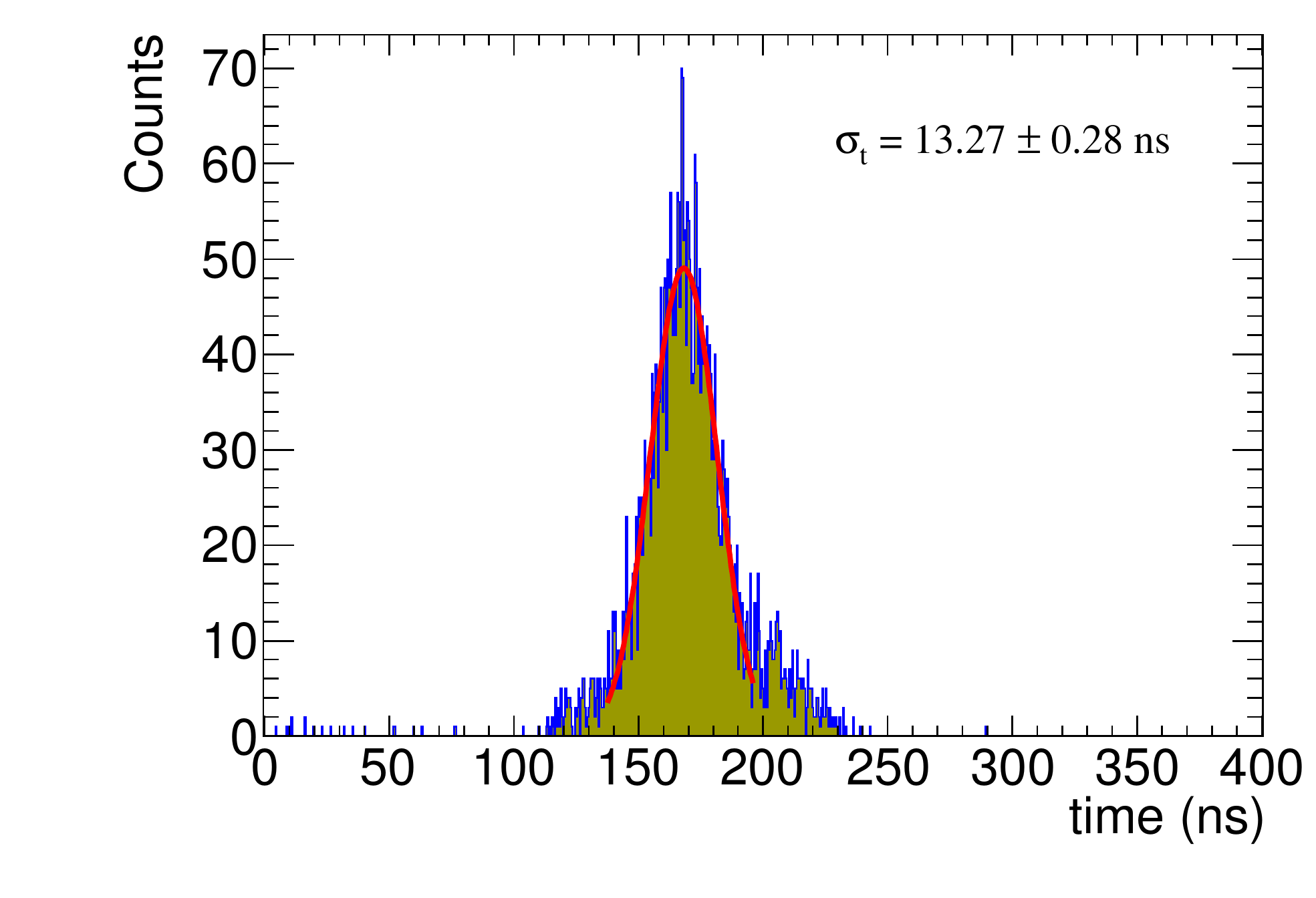}
		\caption{Time spectrum of the quadruple GEM detector.}
		\label{time_spectrum}
	\end{figure}
	
	The time resolution of a detector is the measurement of the spread of the time for a set of events. 
	The time resolution of the quadruple GEM detector is measured in Ar/CO$_{2}$ 70:30 gas mixture. The threefold scintillator trigger setup is used in this measurement with $^{106}$Ru-Rh $\beta^{-}$-source providing the trigger. 
	The detector signal is passed through the timing output of ORTEC 142IH Pre-Amp to a fast amplifier. ORTEC model-454 timing filter amplifier (TFA) is used with integral and differential time both set to 50~ns which is optimized to have a sharp clear signal with minimum noise fluctuation. The threefold trigger signal from the scintillators is used as a start signal and the amplified detector signal after discriminator is used as the stop signal for the ORTEC 567 Time to Amplitude Converter (TAC) module. The time difference between the start and the stop signals provides the time spectrum, which is shown in Fig.~\ref{time_spectrum}.
	
	\begin{figure}[!th]
		\centering
		\includegraphics[width=2.7in, height=2.0in]{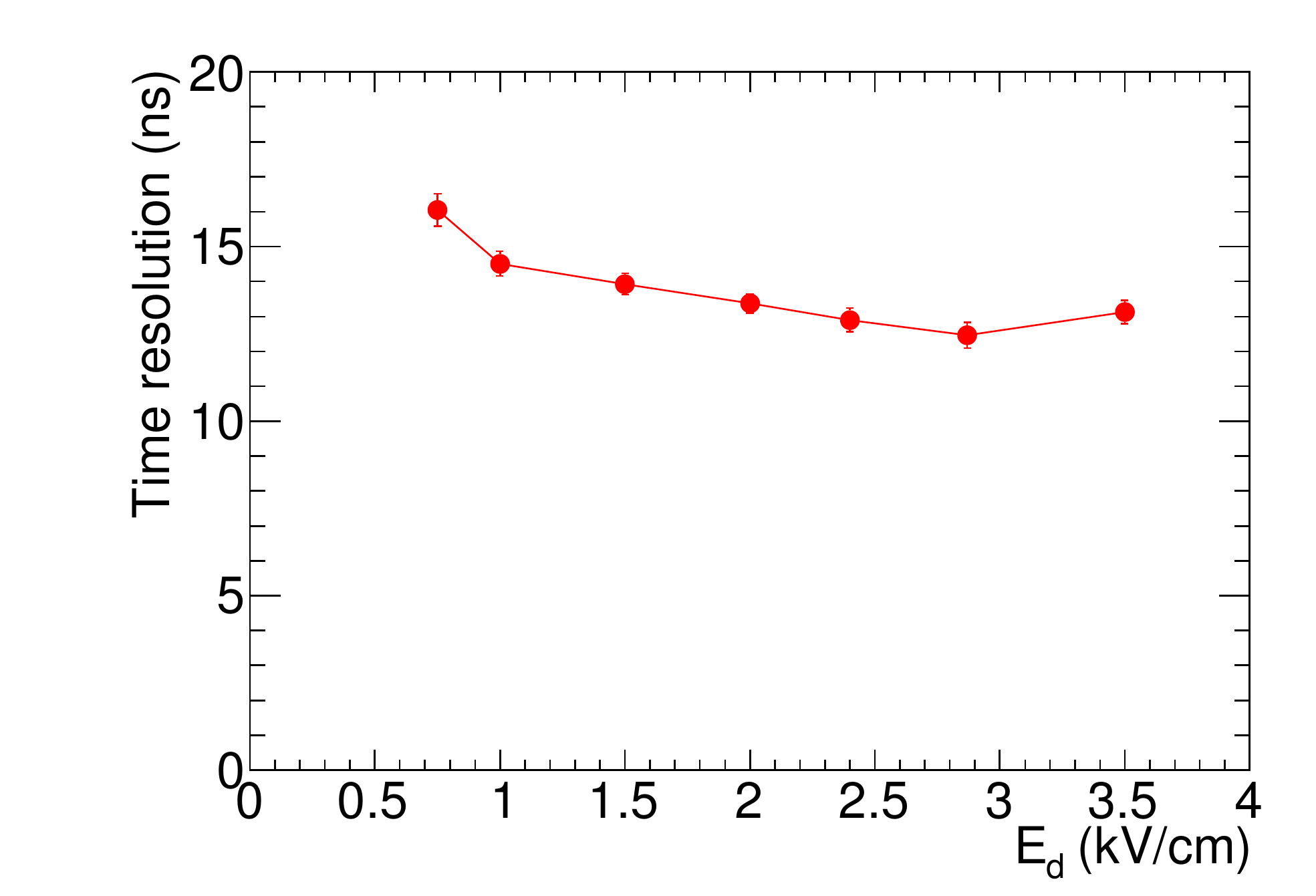}
		\caption{Variation of time resolution as a function of drift field (E$_{d}$) of the quadruple GEM detector. The statistical errors are small and within the symbol size.}
		\label{time_reso_ed}
	\end{figure}
	Time resolution of a detector depends on the property of the gas mixtures,  specifically on the 
	electron drift velocity and thus the drift field.
	The effects of drift field (E$_{d}$) on time resolution has been studied and is shown in Fig.~\ref{time_reso_ed}.
	For this study, the detector gain was set at 
	$\sim$10$^{4}$ and the GEM voltages were also kept constant throughout the measurement.
	It is observed that time resolution decreases with drift field and
	remains almost constant at higher drift field. 
	The time resolution of a gas detector depends on how fast the primary
	electron cluster reaches the readout. This depends on the drift
	velocity of the electrons. In Ar/CO$_{2}$ 70:30 gas mixture, the drift
	velocity increases with the increase of electric field and remains almost
	constant above 2~kV/cm. The time resolution decreases with drift field
	and does not improve much with increasing the drift field above
	2~kV/cm. The optimum value of the time resolution, $\sim$13~ns has been
	obtained with the present setup. In this measurement scintillators
	timing correction is also considered. 
	It is to be noted that in this setup time walk correction of the GEM
	detector could not be taken care of. The time resolution of the
	quadruple GEM detector is slightly worse than the measured value of
	triple GEM~\cite{benci, rajendra2017}.

	\section{Summary}
	
	A quadruple GEM detector has been assembled and tested keeping in mind its usefulness in future HEP experiments. For comparative studies, the detector has been operated
	with Ar/CO$_{2}$ at 90:10 and 70:30 gas mixtures. 
	The detector has been tested with using 
	$^{55}$Fe and $^{106}$Ru-Rh radioactive sources. Basic characteristics like energy spectrum, gain, energy resolution, and efficiency have been measured and compared with that of triple GEM detector setup.
	Usefulness of the quadruple GEM detector is understood in terms of the
	low operating voltage 
	($\Delta V_{GEM-single}$) compared to the triple GEM at similar gain
	responses. Low operating voltages are preferable for long term stable
	operation. However, quadruple GEM detector has somewhat poorer energy resolution compared to the triple GEM, which is consistent with earlier measurements. The time resolution of the quadruple GEM detector is also measured with different E$_{d}$ and the obtained optimum value is $\sim$13~ns.
	We have shown the importance of gain measurements and intricate relationships of gain and efficiency measurements. The influence of the drift field (E$_{d}$) on primary electron transparency has been shown and E$_{d}$ has to be optimized to transfer a maximum number of electrons from the drift volume to the top GEM foil. 
	\bigskip
	\noindent
	\section {Acknowledgment}
	RNP acknowledges the receipt of UGC-NET fellowship. YPV thanks
	Indian National Science Academy, New Delhi for the
	Senior Scientist position. 

\end{document}